==================

\documentstyle[12pt]{article}
\begin{document}

% Useful definitions

\newcommand{\form}[1]{(\ref{#1})}
\def\ojo{\fbox{\bf !`ojo!}}
\def\adhoc{{\it ad hoc}}
\def\etal{{\it et al.}}
\def\eg{{\it e.g.}}
\def\ie{{\it i.e.}}
\def\lsim{\mathrel{\lower2.5pt\vbox{\lineskip=0pt\baselineskip=0pt
           \hbox{$<$}\hbox{$\sim$}}}}
\def\gsim{\mathrel{\lower2.5pt\vbox{\lineskip=0pt\baselineskip=0pt
           \hbox{$>$}\hbox{$\sim$}}}}

% Greek letters

\newcommand{\al}{\alpha}
\newcommand{\bt}{\beta}
\newcommand{\ga}{\gamma}
\newcommand{\Ga}{\Gamma}
\newcommand{\de}{\delta}
\newcommand{\De}{\Delta}
\newcommand{\lam}{\lambda}
\newcommand{\Lam}{\Lambda}
\newcommand{\e}{\epsilon}
\newcommand{\k}{\kappa}
\newcommand{\si}{\sigma}
\newcommand{\Si}{\Sigma}
\newcommand{\om}{\omega}
\newcommand{\Om}{\Omega}
\newcommand{\vphi}{\varphi}

% Mathematical symbols

\newcommand{\lp}{\left(}
\newcommand{\rp}{\right)}
\newcommand{\lll}{\left\{}
\newcommand{\rll}{\right\}}
\newcommand{\lc}{\left[}
\newcommand{\rc}{\right]}
\newcommand{\med}{\frac{1}{2}}
\newcommand{\re}{\mbox{Re}\,}
\newcommand{\im}{\mbox{Im}\,}
\newcommand{\p}{^{\prime}}
\newcommand{\pp}{^{\prime\prime}}
\newcommand{\R}{I\!\! R}
\newcommand{\1}{I\!\! I}
\newcommand{\intoi}{\int_0^{\infty}\!\!\!}
\newcommand{\cH}{ {\cal H}}
\newcommand{\cM}{  {\cal M}}
\newcommand{\cC}{ {\cal C} }
\newcommand{\cd}{ {\cal D}}
\newcommand{\hs}{\hspace{5mm}}
\newcommand{\vs}{\vspace{2mm}}
\newcommand{\Ai}{\mbox{Ai}}
\newcommand{\Bi}{\mbox{Bi}}
\load{\Large}{\sc}\newcommand{\G}{  {\sc g} }
\newcommand{\reg}{{\bigcirc\!\!\!\!\!{\scriptsize R}}}
\newcommand{\tcu}{ {^{^{(3)}}\!\!R} }
\newcommand{\raw}{\rightarrow}
\newcommand{\ket}[1]{\left| #1\right\rangle}
\newcommand{\bra}[1]{\left\langle #1\right|}
\newcommand{\braket}[2]{\left\langle #1\right|\left. #2\right\rangle}
\newcommand{\ab}{{\al\beta}}
\newcommand{\gab}{ g_\ab}
\newcommand{\gsab}{ g^\ab}
\newcommand{\qsik}{ q^{ik}}
\newcommand{\qik}{ q_{ik}}

%DEFinitions

\newcommand{\wh}{wormhole}
\newcommand{\st}{spacetime}
\newcommand{\wf}{wave function}
\newcommand{\wdwe}{Wheeler--DeWitt equation}
\newcommand{\sch}{Schr\"odinger}
\newcommand{\mss}{minisuperspace}
\newcommand{\wdw}{Wheeler--DeWitt}
\newcommand{\tg}{three--geometry}
\newcommand{\tgs}{three--geometries}
\newcommand{\fg}{four--geometry}
\newcommand{\fgs}{four--geometries}
\newcommand{\tm}{three--metric}
\newcommand{\fm}{four--metric}
\newcommand{\ts}{three--surface}
\newcommand{\pin}{path integral}
\newcommand{\ase}{asymptotically Euclidean}
\newcommand{\bc}{boundary condition}
\newcommand{\qc}{quantum cosmology}
\newcommand{\qg}{quantum gravity}
\newcommand{\gr}{general relativity}
\newcommand{\hh}{Hartle--Hawking}
\newcommand{\diff}{diffeomorphism}
\newcommand{\dofs}{degrees of freedom}

%DEF titlepage

\newcommand{\nfc}{{\tt
                  \vspace{.5cm}
                  \begin{center}
                  Preliminary version\\Not for circulation
                  \end{center}}}
\newcommand{\iop}{{\it
                 Instituto de Optica\\
                 Consejo Superior de Investigaciones Cient\'{\i}ficas\\
                 Serrano 121, E--28006 Madrid, Spain}}

%DEFiniciones de ecuaciones

\def\eel#1{\label{#1}\end{equation}}
\def\eeal#1{\label{#1}\end{eqnarray}}
\def\bel#1{\begin{equation}\label{#1}}
\newcommand{\be}{\begin{equation}}
\newcommand{\bea}{\begin{eqnarray}}
\newcommand{\ba}{\begin{array}}
\newcommand{\ee}{\end{equation}}
\newcommand{\eea}{\end{eqnarray}}
\newcommand{\ea}{\end{array}}
\newcommand{\nl}{\nonumber\\}
\newcommand{\ben}{\[}
\newcommand{\een}{\]}

%DEFiniciones de revistas

\newcommand{\bb}{\bibitem}
\newcommand{\rev}[4]{{ \it  #1} {\bf #2}, #3 (#4).}
\newcommand{\npb}[4]{{ \it  #1} {\bf #2} (#3) #4.}
\newcommand{\np}{Nucl. Phys.}
\newcommand{\pr}{Phys. Rev.}
\newcommand{\cqg}{Class. Quant. Grav.}
\newcommand{\pl}{Phys. Lett.}
\newcommand{\prl}{Phys. Rev. Lett.}
\newcommand{\prep}{Phys. Rep.}
\newcommand{\apj}{Astrophys. J.}
\newcommand{\ijmp}{Int. J. Mod. Phys.}
\newcommand{\mpl}{Mod. Phys. Lett.}
\newcommand{\fp}{Found. Phys.}
\newcommand{\jmp}{J. Math. Phys.}
\newcommand{\cmp}{Commun. Math. Phys.}
\newcommand{\rpp}{Rep. Prog. Phys.}

%DEFiniciones en castellano

\newcommand{\fo}{funci\'on de onda}
\newcommand{\fos}{funciones de onda}
\newcommand{\uni}{universo}
\newcommand{\topo}{topolog\'{\i}a}
\newcommand{\cet}{cambios en la \topo}
\newcommand{\ct}{cambios topol\'ogicos}
\newcommand{\ag}{agujero de gusano}
\newcommand{\an}{agujero negro}
\newcommand{\ags}{agujeros de gusano}
\newcommand{\ans}{agujeros negros}
\newcommand{\et}{espaciotiempo}
\newcommand{\ic}{integral de camino}
\newcommand{\ub}{\uni\ beb\'e}
\newcommand{\ubs}{\uni s beb\'e}
\newcommand{\etl}{espaciotemporal}
\newcommand{\cq}{cosmolog\'{\i}a cu\'antica}
\newcommand{\ccs}{condiciones de  contorno}
\newcommand{\cc}{condici\'on de contorno}
\newcommand{\rg}{relatividad general}
\newcommand{\gq}{gravedad cu\'antica}
\newcommand{\bkg}{backgound?'?}
\newcommand{\tqc}{teor\'{\i}a cu\ ntica de campos}
\newcommand{\ics}{integrales de camino}
\newcommand{\mq}{mec\'anica cu\'antica}
\newcommand{\dif}{difeomorfismo}
\newcommand{\tsu}{tres--superficie}
\newcommand{\tme}{tres--m\'etrica}
\newcommand{\rep}{reparametrizaci\'on}
\newcommand{\reps}{reparametrizaciones}
\newcommand{\tge}{tres--geometr\'{\i}a}
\newcommand{\cme}{cuatro--m\'etrica}
\newcommand{\mse}{minisuperespacio}
\newcommand{\gdl}{grado de libertad}
\newcommand{\gdls}{grados de libertad}
\newcommand{\cge}{cuatro--geometr\'{\i}a}
\newcommand{\cva}{cuatro--variedad}
\newcommand{\asp}{asint\'oticamente plan}

\newcommand{\nocast}[1]{{\it #1}}

%Definiciones auxiliares #tiles

\newcommand{\figaux}[1]{\newline
                        \begin{center}\ojo\ \hs Figura #1. \end{center}}

\newcommand{\nsection}[1]{  \begin{center}
			    \section{#1}
			    \end{center}
			    \indent\setcounter{equation}{0}}
\newcommand{\nsubsection}[1]{\begin{center}
			    \subsection{#1}
			    \end{center}
			    \indent}
\newcommand{\nsubsubsection}[1]{\subsubsection{#1}\indent}
\newcommand{\nnsection}[1]{\begin{center}
			    \section*{#1}
			    \end{center}
			    \indent\setcounter{equation}{0}}
\newcommand{\nnsubsection}[1]{\begin{center}
			    \subsection*{#1}
			    \end{center}
			    \indent}
\newcommand{\nnsubsubsection}[1]{\subsubsection*{#1}\indent}
\renewcommand{\thesection}{\arabic{section}.}
\renewcommand{\thesubsection}{\thesection\arabic{subsection}.}
\renewcommand{\thesubsubsection}{\thesubsection\arabic{subsubsection}.}
\renewcommand{\theequation}{\arabic{section}.\arabic{equation}}

\begin{titlepage}

\title{{\bf Jordan--Brans--Dicke Quantum Wormholes \\
and Coleman's Mechanism}
\thanks{Work partially supported by CICYT under
contract PB91-0052 and AEN90--0139.}}

\author{
{\bf Luis J. Garay}\thanks{Supported by
MEC--FPI Grant. e--mail: garay@cc.csic.es} \\
Instituto de Optica, \ CSIC\\
Serrano, 121\ \ \ E--28006\ \ Madrid ,\ \ Spain \vspace{2mm}\\
{\bf Juan Garc\'{\i}a--Bellido}\thanks{Supported by
MEC--FPI Grant. e--mail: bellido@cc.csic.es} \\
Instituto de Estructura de la Materia , \ CSIC\\
Serrano, 123\ \ \ E--28006\ \ Madrid ,\ \ Spain }

\date{}
\maketitle
\def\baselinestretch{1.15}
\begin{abstract}
We consider the quantum gravity and cosmology of a
Jordan--Brans--Dicke theory, predicted by string effective actions.
We study its canonical formalism and find that the constraint algebra
is that of \gr, as a consequence of the general covariance of
scalar--tensor theories. We also analyze the problem of boundary
conditions and propose that they must be imposed in the Jordan frame,
in which particles satisfy the strong equivalence principle.
Spe\-ci\-fi\-cally, we discuss both Hartle--Hawking and \wh\ boundary
conditions in the context of quantum cosmology.  We find quantum
wormhole solutions for Jordan--Brans--Dicke gravity even in the
absence of matter. Wormholes may affect the constants of nature and,
in particular, the Brans--Dicke parameter. Following Coleman's
mechanism, we find a probability distribution which is strongly
peaked at zero cosmological constant and infinite Brans--Dicke
parameter.  That is, we recover \gr\ as the effective low energy
theory of gravity.

\end{abstract}

%\vspace{-5mm}\nfc

\vskip-19cm
\rightline{\bf IEM--FT--59/92}
\rightline{\bf October 1992}
\vskip3in

\end{titlepage}

\newpage
\def\baselinestretch{1.25}

\nsection{Introduction}

It is generally believed that the theory of gravity at low energies
(general relativity, scalar--tensor, etc.) must be an effective
approximation of a more fundamental theory of quantum gravity.
Nowadays, the only consistent candidate for such a fundamental theory
is string theory \cite{GSW}, based on the assumption that the
fundamental objects that describe matter and its interactions are not
point--like but one--dimensional. This simple assumption gives a
consistent description of gravity at energies above the Planck scale,
$M_{\rm Pl}$, and has very interesting consequences. In particular,
the gravitational sector of closed strings contains, apart from the
graviton and the antisymmetric tensor field, a dilaton scalar field
that couples to gravity and matter. The axion, related to the
antisymmetric tensor field, has already been considered in the
context of \qg\ by Giddings and Strominger \cite{GIS}. In this paper
we study the scalar component of the gravitational sector in the same
context.

The low energy string effective theory can be obtained by integrating
out all the string quantum fluctuations with respect to its center of
mass \cite{FTS}, giving a local field theory which has the form of a
scalar--tensor theory of gravity. We can write the tree level
string effective action in four dimensions, keeping only linear terms
in the string tension $\alpha'$ and in the curvature $R$, as
\begin{equation}
\label{SEA}
\tilde{S} = \frac{1}{\alpha'} \int d^{4}x \sqrt{-g}\ e^{-2\phi}
\left(R + 4g^{\alpha\beta}\nabla_\alpha \phi
\nabla_\beta \phi \right) \ + S_m,
\end{equation}
where $\phi$ is the dilaton field. This action is equivalent to a
Jordan--Brans--Dicke (JBD) theory of gravity \cite{JBD},
\begin{equation}
\label{JBD}
\tilde{S} = \frac{1}{16\pi \G} \int d^4x \sqrt{-g} \left(\Phi R
- \frac{\omega}{\Phi} g^{\alpha\beta}\nabla_\alpha\Phi
\nabla_\beta\Phi \right) \ + S_m,
\end{equation}
with a constant Brans--Dicke (BD) parameter, $\omega=-1$. This
particular value is due to a fundamental symmetry of strings: target
space duality \cite{DUA}. This is a symmetry of string amplitudes
which relates large and small radius of compactification. Therefore,
$\omega=-1$ is a model--independent prediction of string theories
\cite{NPB}. One should try to verify phenomenologically this
prediction \cite{CQG,GB92}. Jordan--Brans--Dicke theory is based on
the idea of Mach that inertia arises from accelerations with respect
to the general distribution of matter in the universe. Therefore, the
inertial masses of elementary particles are not fundamental constants
but represent the interaction of particles with a cosmic scalar field
whose dynamics depend on the rest of the matter in the universe.
Supposing that all matter particles have the same coupling to the
scalar field, $m(\phi) = e^{\beta\phi} m$, the action in the so
called Einstein frame can be written as
\begin{equation}
\label{SBD}
\tilde{S} = \frac{1}{16\pi \G} \int d^4x \sqrt{-\bar{g}}
\left(\bar{R} - \frac{1}{2} \bar{g}^{\alpha\beta}
\bar{\nabla}_\alpha\phi
\bar{\nabla}_\beta\phi \right)\ + \int e^{\bt\phi} m\ d\bar{s}.
\end{equation}
Physics must be invariant under conformal redefinitions of the metric
since they correspond to an arbitrary choice of measuring units. We
are thus free to choose the conformal frame in which we want to
describe physics. The most natural choice is the so called physical
frame \cite{CQG}, in which observable particles have constant masses,
since in this frame particles follow geodesics of the metric and thus
satisfy the strong equivalence principle \cite{WEI}.  From the
cosmological point of view, the comoving frame, in which a
fundamental observer sees the universe as homogeneous and isotropic,
is the physical frame since those observers will follow geodesics of
the metric.  In our case, the conformal redefinition of the metric
that allows us to describe the theory of gravitation (\ref{SBD}) in
the physical frame is $\bar{g}_{\alpha\beta} = e^{-2\beta\phi}
g_{\alpha\beta} \equiv \Phi g_{\alpha\beta}$.  This is the so called
Jordan frame, in which the action takes the form \form{JBD} where
$\Phi$ is a dimensionless scalar field and $\omega$ a constant given by
\begin{equation}
\label{OME}
2\omega+3 = \frac{1}{4\beta^2}.
\end{equation}
General relativity is recovered in the limit $\omega =
\infty, \ (\beta = 0)$. On the other hand,
the physical frame for \gr\ is the Einstein frame, in which the
gravitational action takes the usual Hilbert--Einstein form.

These theories are not ruled out by post--Newtonian experiments
\cite{SCT} nor by primordial nucleosynthesis bounds \cite{PNS},
and in fact have recently recovered great interest since they have
been proposed as the arena for extended inflation \cite{EI,EIS}, a
new inflationary scenario which could solve some of the traditional
problems of previous schemes, such as the graceful exit problem of
old inflation \cite{OLD} and the fine tuning problems of new
inflation \cite{NEW}.
Furthermore, as we have mentioned, JBD theory may well be the Planck
scale theory of gravity.

In this paper we will analyze the Jordan--Brans--Dicke quantum
cosmology with special emphasis made on the effect of quantum
wormholes on the low energy coupling constants.\footnote{Since the
fundamental description of strings gives a JBD theory of gravity
below the Planck scale, one should study JBD quantum cosmology, see
\cite{VDC,DCA}, which may be relevant to the problem of initial
conditions for extended inflation. A different approach is string
quantum cosmology, see \cite{SQC}, in which the compactification
ansatz and the dimension of spacetime are analyzed.} In section 2, we
study the canonical formalism of a scalar--tensor theory of gravity.
The classical action \form{JBD} contains second derivatives of the
metric tensor, which may be removed with the addition of a suitable
surface term \cite{H79}. The Hamiltonian constraints can be
explicitly written in both the Jordan and Einstein frames. Both
expressions are related through a canonical transformation. The
constraint algebra is the same in both frames and
corresponds to that of \gr, expressing the fact that JBD theory is
generally covariant. Note that a scalar component of gravity does not
spoil this covariance.

Boundary conditions must be imposed in the physical frame, where
particles satisfy the strong equivalence principle. The boundary
conditions in any other frame can be obtained by transforming those
in the physical frame through the corresponding conformal
redefinition. In JBD quantum cosmology we shall be concerned with
boundary conditions both for the universe \cite{HAL} and for wormholes
\cite{HP}.
The \hh\ \wf\ for the universe \cite{HH} is given by the Euclidean
\pin\ over all compact \fm s in the Jordan frame, over all possible
BD field configurations and all matter fields. We find that the
saddle point configuration, which gives the dominant contribution to
the \pin, will be compact when both the cosmological constant and the
BD parameter are positive.

Non--perturbative quantum gravity effects due to non--trivial
topologies, \eg\ wormholes \cite{BUS}, will change the effective
value of the physical parameters of the theory, in particular the
Brans--Dicke parameter.  Wormholes may be interpreted,
semiclassically, as throats joining two otherwise disconnected large
regions of spacetime. In this picture one assumes the dilute wormhole
approximation, in which wormhole ends are far away from each other
and thus can be treated separately.  JBD wormhole wave functions\linebreak
can be written as Euclidean \pin s over asymptotically flat
spacetimes \cite{HP}. One must also sum over all JBD field
configurations whose Hamiltonian vanishes in the asymptotic region
and similarly for the matter fields \cite{G91}.  These boundary
conditions\linebreak have been imposed in the Jordan frame, although
in the Einstein frame they take the same form, since the JBD field is
constant at infinity. The problem of boundary conditions will be
analyzed in detail in section 3.

Wormholes may play an important role in solving problems associated
with the complete evaporation of black holes \cite{H88}. Furthermore,
although Planck scale wormholes are not directly observable, they
will produce effective interactions in the low energy physics
\cite{H88} that turn the coupling constants of nature into dynamical
variables \cite{HL,C88}. In particular, JBD wormholes will affect the
kinetic term of the JBD field as well as introduce a coupling between
this field and particle masses. Both effects will modify the value of
the low energy BD parameter.  It is worth noticing that there exist
wormhole solutions even in the absence of matter, in contrast with
the situation in general relativity.  Wormholes in JBD gravity are
discussed in section 4.

Wormholes not only affect the constants of nature but also may
provide a probability distribution for them. In \gr\ this
distribution is strongly peaked at zero cosmological constant.  This
is the so called Coleman's mechanism \cite{COL} for the vanishing of
the cosmological constant. In JBD theory an
analogous argument leads to the conclusion that not only the
cosmological constant vanishes but also the BD para\-me\-ter is driven to
infinity, and consequently there cannot be an effective dynamical scalar
field coupled to any form of energy or matter.
Thus we recover \gr\ as the low energy effective theory
of gravity, even though the fundamental high energy description may
well be a scalar--tensor theory, as suggested by strings.
This result precludes\linebreak {\it extended} inflation,
since the predicted values for the cosmological constant and the BD
parameter are the low energy effective values at all times \cite{COL}.
Section 5 is devoted to Coleman's mechanism in JBD theory.

\nsection{Canonical formalism}

The Euclidean action for the Jordan--Brans--Dicke theory of
gravity, in the Jordan frame, can be obtained through a Wick
rotation ($t\rightarrow-i\tau$) from \form{JBD}
\be
\tilde{I}=\frac{1}{16\pi\G}\int d^4x\sqrt g\lll-\Phi R+\frac{\om}{\Phi}
g^{\al\beta}\nabla_\al\Phi\nabla_\beta\Phi\rll+I_m[g_{\al\beta},
\sigma],
\eel{IBD}
where $I_m[g_{\al\beta},\sigma]$ is the Euclidean action of the
matter fields $\sigma$.  In a (3+1) slicing of \st, the metric can be
written in the standard from \cite{ADM,MTW}
\be
ds^2=(N^2+N_iN^i)d\tau^2+2N_idx^id\tau+q_{ik}dx^idx^k,
\eel{DS2}
where $N$ is the lapse function which measures the proper time
separation between two neighbouring three--sections, $N^i$ are
the shift functions which measure the separation between the
lines of constant $x^i$ and the normal to the \ts\ $\Si_\tau$,
and $q_{ik}$ is the metric on this \ts, which will be chosen to be
connected, compact and without boundary. Then, the Euclidean
action takes the form
\be\ba{rl}
\tilde{I}\ =&{\displaystyle \frac{1}{16\pi\G}\int d^3x\ d\tau N\sqrt
q\lll\Phi(K_{ik}K^{ik}-K^2)-\Phi \tcu\right.}\vs\nl
+&{\displaystyle \left.\frac{\om}{\Phi}(D\Phi)^2+\frac{\om}{\Phi}q^{ik}
\nabla_i\Phi\nabla_k\Phi-2\Phi\nabla_\al f^\al\rll+
I_m[g_{\al\beta},\si], }
\ea\eel{I3D}
where  $q=\det q_{ik}$,
$D\equiv n^\al\nabla_\al$ is the derivative in the normal direction
to the \ts, $\displaystyle n^\al=\lp\frac{1}{N},-\frac{N^i}{N}\rp$,
and
\be
f^\al=Kn^\al+Dn^\al.
\eel{FKD}
$K$ is the trace of its second fundamental
form,
\be
K_{ik}=\frac{1}{2N}\lp-\dot{q}_{ik}+2\nabla_{(i}N_{k)}\rp,
\eel{KIK}
where an overdot denotes a partial derivative with respect to $\tau$.
The expression $\nabla_\al f^\al$ contains second time derivatives of
the metric. Therefore, the last term in the gravitational Euclidean
action can be integrated by parts to give a surface term plus a
series of terms which contain only first time derivatives of the
three--metric and the JBD field
\be\ba{rl}
\tilde I \ =&{\displaystyle \frac{1}{16\pi\G}\int d^3x d\tau N\sqrt q\lll\Phi
(K_{ik}K^{ik}-K^2)-\Phi\tcu \right.}\vs\nl +&{\displaystyle \left.
\frac{\om}{\Phi}(D\Phi)^2+2KD\Phi+\om q^{ik}\nabla_i\Phi\nabla_k\Phi+
2q^{ik}\nabla_i\nabla_k\Phi\rll}\vs\nl
+&{\displaystyle I_m[{q}_{ik},N,N^i,\si]-
\frac{1}{8\pi\G}\int d^3x\sqrt q \Phi K. }
\ea\eel{IRK}
In order to be consistent with boundary conditions that fix the \tm,
the JBD and the matter fields at two given \ts s \cite{H79}, we must
remove the surface term
\be
I=\tilde I+\frac{1}{8\pi\G}\int d^3x\sqrt q \Phi K.
\eel{ACT1}
Otherwise, we would have to impose the additional requirement that
the normal derivatives of the \tm\ should also be kept fixed at the
boundaries.  The Einstein--Jordan--Brans--Dicke and the matter field
equations can then be obtained by requiring that the Euclidean action
\form{ACT1} be stationary  under variations of the fields subjected
to the previous boundary conditions.

For the sake of definiteness, we consider a minimally coupled scalar
field $\si$ as  matter content whose action will be
\be
I_m[q_{ik},N,N^i\si]=\int d^3xd\tau  N\sqrt q\lll\med (D\si)^2+\med q^{ik}
\nabla_i\si\nabla_k\si+V(\si)\rll.
\eel{ACM1}
  The canonical momenta are given by
\be\ba{rl}
p^{ik}\ =&{\displaystyle \frac{\de I}{\de\dot q_{ik}}=
\frac{1}{16\pi\G}\sqrt q\lll \Phi(-K^{ik}+q^{ik}K)-
q^{ik}D\Phi\rll,}\vs\nl
p_\Phi\ =&{\displaystyle \frac{\de I}{\de\dot \Phi}=
\frac{1}{8\pi\G}\sqrt q\ \frac{\om}{\Phi}D\Phi,}\vs\nl
p_\si\ =&{\displaystyle \frac{\de I}{\de\dot\si}=\sqrt q D\si }
\ea\eel{MOM1}
and, therefore, the Hamiltonian of the theory is
\be
H=\int d^3x \lp N\cH+   N^i \cH_i\rp,
\eel{HNH}
where the Hamiltonian generators acquire the quite involved
expressions
\be\ba{rl}
\cH\ =&{\displaystyle 16\pi\G\frac{1}{\Phi\sqrt q}\lll p^{ij}p^{kl}\lp
q_{ik}q_{jl}-\frac{\om+1}{2\om+3}q_{ij}q_{kl}\rp\right.}\vs\nl
-&{\displaystyle \left.\frac{1}{2\om+3}\Phi p_\Phi p +
\frac{1}{2(2\om+3)}\Phi^2p_\Phi^2\rll}\vs\nl
+&{\displaystyle \frac{1}{16\pi\G}\sqrt q\lll\Phi\tcu-
\om q^{ik}\nabla_i\Phi\nabla_k\Phi\rll}\vs\nl
+&{\displaystyle  \frac{1}{2\sqrt q}p_\si^2-\med\sqrt
qq^{ik}\nabla_i\si\nabla_k\si-\sqrt q V(\si), }\vspace{3mm}\nl
\cH_i\ =&{\displaystyle -2q_{ik}\nabla_jp^{jk}+p_\Phi\nabla_i\Phi+
p_\si\nabla_i\si. }
\ea\eel{HIH}

The constraints of the theory, which together represent the
invariance under spatial diffeomorphisms and time
reparametrizations, are
\be
\cH=0,
\hs
  \cH_i=0.
\eel{HHI}
It can be seen that the Poisson bracket algebra of the constraints is
that of \gr\ although the direct computation of the constraint algebra
is a little messy. However, a simple argument will allow us to obtain
this result in a different way. As mentioned in the introduction, due
to the equivalence under conformal redefinitions, we can work in the
Einstein frame in which the gravitational coupling becomes constant
while the matter fields are non--trivially coupled to the JBD scalar.
Since this is a non derivative coupling, the canonical structure of
the theory will not change.  More explicitly, the canonical
transformation
\be
\bar{q}_{ik}=\Phi q_{ik},
\hs
\phi = - \frac{1}{2\beta}\log\Phi,
\eel{CT}
together with the redefinitions  of the lapse and shift functions
\be
\bar{ N}^2=\Phi N^2,
\hs
\bar{N}_i=\Phi N_i,
\eel{DEF}
gives the action in the Einstein frame
\be\ba{rl}
\bar{ I}\ =&{\displaystyle \frac{1}{16\pi\G}\int\!\! d^3xd\tau
\bar N\sqrt{ \bar{q}}\lll\bar K_{ik}\bar K^{ik}-
\bar K^2-{^{^{(3)}}\!\!\bar R}+\med(\bar D\phi)^2+\med
\bar{q}^{ik}\bar{\nabla}_i\phi\bar{\nabla}_k\phi\rll}\vs\nl
+&{\displaystyle \bar I_m[\bar{q}_{ik},\bar N,\bar N^i,\si,\phi]. }
\ea\eel{ACT2}
Note that the matter action
\be
\bar I_m[\bar{q}_{ik},\bar N,\bar N^i,\si,\phi]=\int d^3xd\tau \bar N
\sqrt{ \bar{q}} e^{2\beta\phi} \lll\med(\bar D\si)^2
+\med \bar{q}^{ik}\bar{\nabla}_i\si\bar{\nabla}_k\si+
e^{2\beta\phi}V(\si)\rll
\eel{ACM2}
has now an explicit dependence on $\phi$.

The canonical momenta conjugate to the canonical variables
$\bar{q}_{ik}$, $\phi$ and $\si$ are
\be\ba{rl}
\bar p^{ik}\ =&{\displaystyle -\frac{1}{16\pi\G}\sqrt{ \bar{q}}(\bar
K^{ik}-\bar{q}^{ik}\bar K),}\vs\nl
\bar p_\phi\ =&{\displaystyle \frac{1}{16\pi\G}\sqrt{ \bar{q}}\bar
D\phi,}\vs\nl
\bar p_{\si}\ =&{\displaystyle \sqrt{ \bar{q}}e^{2\beta\phi}D\si,}
\ea\eel{MOM2}
which can be obtained either directly from the action
\form{ACT2} or from \form{MOM1} by means of the canonical
transformation \form{CT}.  The Euclidean action in the Einstein
frame can then be written  in terms of the canonical variables
and momenta as
\be
\bar I=\int d^3xd\tau(\bar p^{ik}\dot{ \bar{q}}_{ik}
+\bar p_\phi\dot \phi+\bar p_\si\dot \si-
\bar N\bar\cH-\bar N^i\bar\cH_i),
\eel{INH}
where the Hamiltonian generators $\bar\cH$ and $\bar\cH_i$ take
the form
\be\ba{rl}
\bar\cH\ =&{\displaystyle 16\pi\G \frac{1}{\sqrt{ \bar{q}}}
\lp \bar p^{ik}\bar p_{ik} -\med \bar p^2+\med\bar p_\phi^2\rp}\vs\nl
+&{\displaystyle \frac{1}{16\pi\G}\sqrt{ \bar{q}}\lp {^{^{(3)}}\!\!\bar R}
-\bar{q}^{ik}\bar{\nabla}_i\phi\bar{\nabla}_k\phi \rp}\vs\nl
+&{\displaystyle  e^{-2\beta\phi}\frac{1}{2\sqrt{
\bar{q}}}\bar p_\si^2-\med e^{2\beta\phi}\sqrt{ \bar{q}}
\bar{q}^{ik}\bar{\nabla}_i\si\bar{\nabla}_k\si- e^{4\beta\phi}
\sqrt{\bar{q}}V(\si), }\vspace{3mm}\nl
\bar\cH_i\ =&{\displaystyle -2\bar{q}_{ij}\bar{\nabla}_k\bar p^{jk}+
\bar p_\phi\bar{\nabla}_i\phi+\bar p_\si\bar{\nabla}_i\si,}
\ea\eel{BHH}
in terms of which the Hamiltonian of the theory can be written,
as before, as
\be
\bar H=\int\!\! d^3x\lp \bar N\bar\cH+ \bar N^i\bar\cH_i\rp.
\eel{BNH}
It is straightforward to see that the Poisson bracket algebra of
the constraints $\bar \cH$ and $\bar \cH_i$ is that of \gr\ \cite{KUC}
\be\ba{rl}
{\displaystyle
\lll \int\!\! d^3x\bar N(x)\bar\cH(x),\int \!\!d^3y\bar
M(y)\bar\cH(y)\rll}\ =&{\displaystyle
-\int\!\!d^3z\lp\bar N(z)\stackrel{\leftrightarrow}{\bar \nabla}\hspace{0cm}^i
\bar M(z)\rp\bar\cH_i(z),}\vs\nl
{\displaystyle
\lll \int\!\!d^3x \bar N^i(x)\bar\cH_i(x), \int\!\!d^3y
\bar{ M}^k(y)\bar\cH_k(y)\rll}\ =&{\displaystyle
-\int\!\!d^3z\lp \bar{\cal L}_{{\vec N}}\bar{ M}^k(z)\rp\bar\cH_k(z),}\vs\nl
{\displaystyle \lll \int\!\!d^3x\bar N\bar\cH(x),
\int\!\!d^3y\bar{M}^i(y)\bar\cH_i(y)\rll}\ =&{\displaystyle
-\int\!\!d^3z \lp \bar{\cal L}_{{\vec M}}\bar{N}(z)\rp\bar\cH(z),  }
\ea\eel{PBA}
where $\bar{\cal L}_{\vec v}$ is the Lie derivative along the vector
$ \bar v^i$.  Since the first relation involves the canonical
variable $\bar{q}_{ik}$, \ie\ $\bar \nabla^i=\bar{q}^{ik}\bar{\nabla}_k$,
this algebra is not the Lie algebra of \st\ diffeomorphisms
\cite{ISK}. This was expected since
Jordan--Brans--Dicke theory is generally covariant and we have
performed a (3+1) splitting of \st, as required by the ADM
formalism, which does not preserve the group structure of the
four diffeomorphisms.

The general theory of canonical transformations ensures
that, for the transformation \form{CT}, both
Hamiltonians must be equal \cite{GOL},
\be
H=\int\!\!d^3x\lp N\cH+N^i\cH_i\rp=\int\!\!d^3x\lp\bar
N\bar\cH+\bar N^i\bar\cH_i\rp
=\bar H.
\eel{HBH}
Since $N$ and $ N^i$ are just Lagrange
multipliers and using \form{DEF}, we find that
\be
\cH=e^{-\beta\phi}\bar\cH,
\hs\hs
\cH_i=\bar\cH_i,
\eel{HHH}
so that
\be
\int\!\!d^3x N\cH=\int\!\!d^3x\bar N\bar\cH,
\hs\hs
\int\!\!d^3x N^i\cH_i=\int\!\!d^3x\bar N^i\bar\cH_i.
\eel{HNI}
Finally, the Poisson brackets are also invariant under canonical
transformations and, therefore, the algebra of constraints will have
the same form \form{PBA} in both frames. This result was also expected
since the change from one frame to the other must not affect the local
\st\ structure nor the slicing procedure.  It is just a redefinition
of the variables on each \ts\ without any reference to their
embeddings in the  \st\ manifold.

\nsection{Boundary conditions}

In the preceeding section we studied the classical equivalence under
canonical transformations of the JBD local dynamical laws, between
the Jordan and the Einstein frames. These canonical transformations
are in fact conformal redefinitions of the metric.  Physics must be
invariant under conformal redefinitions of the metric since they
correspond to an arbitrary choice of measuring units: physical
observables must be constructed as dimensionless variables
\cite{JBD}. In particular, any distance measured in Compton
wavelengths of observable particles is invariant under conformal
redefinitions \cite{CQG}. We can therefore arbitrarily choose the
conformal frame in which we want to des\-cribe physics \cite{CMP}.
However, to describe physical phenomena one not only needs local laws
but also boundary conditions that represent global features of such
phenomena.  Therefore, two physical theories will be equivalent under
conformal redefinitions only if we transform the boundary conditions
as well as the local laws. In this section we analyze the boundary
conditions for the universe and those for wormholes in the context of
Jordan--Brans--Dicke \qc.

The wave function of the universe will be a solution of the \wdwe\
and the quantum version of the \diff\ constraints \form{HHI} \cite{KUC}
\be
\cH\Psi=0,
\hs\hs
\cH_i\Psi=0,
\eel{QHH}
in which the classical variables and momenta become operators which
act on the \wf\ by means of the Euclidean correspondence principle.
In particular, the momenta can be written as functional derivatives
with respect to the canonical variables.  The \wf\ must also satisfy
appropriate boundary conditions. We shall be particularly concerned
with boundary conditions of the \hh\ type \cite{HH}
\footnote{Tunneling boundary conditions for the universe in JBD \qc\
have recently been discussed by Vilenkin and del Campo
\cite{VDC,DCA}. In particular, they study the initial conditions for
extended inflation.}, since we are interested in Coleman's mechanism
\cite{COL} in the JBD theory, as will be analyzed in section 5.  The
\hh\ boundary conditions in a JBD theory of gravity can be stated as
follows: the wave function of the universe is given by the Euclidean
\pin\ over all compact manifolds and all gravitational and matter
fields defined on them which match the arguments of the \wf.  As
mentioned in the introduction, fundamental observers are in the
Jordan frame and therefore the boundary conditions for the universe
must be imposed in this physical frame.  The boundary conditions in
the Einstein frame will be obtained by transforming these in the same
way that one transforms the local laws.

For simplicity, let us consider the case of JBD pure gravity plus a
cosmological constant. The Euclidean action can then be written as
\be
\tilde{I}=\frac{1}{16\pi\G}\int d^4x\sqrt g\lll-\Phi R+\frac{\om}{\Phi}
g^{\al\beta}\nabla_\al\Phi\nabla_\beta\Phi + 2\Lam\rll,
\eel{EBD}
where $\Lam$ is the cosmological constant.\footnote{In the Einstein
frame the cosmological term acquires a $\Phi$ dependence given by
$\displaystyle \frac{2\Lam}{\Phi^2}$.} Note that in the Jordan frame
the cosmological term is in fact constant and thus satisfies the
strong equivalence principle (see also \cite{SCT}).
The saddle point of the Euclidean action, which gives maximum
contribution to the path integral, is a solution of the classical
Euclidean equations of motion
\begin{equation}
\label{SPF}
\begin{array}{c}
{\displaystyle
R_\ab-\med \gab R= \frac{\Lam}{\Phi} \gab + \frac{\om}{\Phi^2}
\lp\nabla_\al\Phi\nabla_\beta\Phi - \med \gab (\nabla\Phi)^2\rp
+ \frac{1}{\Phi}\lp\nabla_\al\nabla_\beta\Phi - \gab \nabla^2\Phi
\rp ,}\vspace{2mm}\\
{\displaystyle
\nabla^2\Phi = \frac{4\Lam}{2\om+3} }.
\end{array}
\end{equation}
The curvature scalar at the saddle point has the form
\be
R = \frac{2\om}{2\om+3}\ \frac{4\Lam}{\Phi} + \om
\lp\frac{\nabla \Phi}{\Phi}\rp^2.
\eel{R4}
It will correspond to a closed universe only when both the
cosmological constant $\Lam$ and the BD parameter $\om$ are
positive. Otherwise, the saddle point configurations are non--compact
and therefore are not considered in the path integral. Furthermore, a
positive cosmological constant damps the wave function for large \fgs,
while a negative one enhances them \cite{HH}, and similarly for the
BD parameter. We shall therefore consider the
case in which $\Lam>0$ and $\om>0$.

In the Einstein frame, the saddle point will be compact even for
$\om<0$, since the curvature scalar acquires the form\footnote{Note
that $2\om+3$ must be positive in order to ensure that gravity
is attractive.}
\be
\bar R = \frac{4\Lam}{\Phi^2} + \frac{2\om+3}{2}
\lp\frac{\bar{\nabla} \Phi}{\Phi}\rp^2.
\eel{BR4}
In this frame, the vacuum energy $\displaystyle \int
d^3 x \sqrt{\bar q(x,t)} \frac{2\Lam}{\Phi(x,t)^2} $ becomes infinite
for negative values of the BD parameter, when
the scalar field vanishes. In fact, these spatial
surfaces correspond to non--compact \ts s in the Jordan frame.  On
the other hand, for $\om>0$ the saddle point will be compact in both
frames and the vacuum energy finite. This supports the statement
that the Jordan frame is the physical frame in which the boundary conditions
must be imposed, as stressed above. In fact, imposing \hh\ boundary conditions
in
the Einstein frame leads to inconsistencies.

The \wh\ \wf\ will be the solution of both the \wdwe\ and the quantum
\diff\ constraints \form{QHH} subjected to suitable \wh\ boundary
conditions in the Jordan frame.  Since  the \st\ is \ase, the \wf\
must be exponentially damped for large \tgs\ \cite{HP}.  It  must also
be regular \cite{HP,G92} when the \tg\ degenerates, since the \fg\ is
non--singular.  The \wh\ \wf\ can also be written as a path integral
over  all \fgs\ which are \ase\ and over all possible configurations
of the JBD and the matter fields which do not present asymptotic
excitations. This means that the gravitational, the JBD and the
matter parts of the Hamiltonian must vanish separately in the
asymptotic region and, in particular, the JBD field must acquire there
a constant homogeneous configuration \cite{G91}.  Therefore, in the
Einstein frame these boundary conditions are expressed in the same
form as those in the Jordan frame.  Since, in the Einstein frame, the
Hamiltonian constraints have a simpler form, the quantization
procedure will be carried out in this frame.

\nsection{Jordan--Brans--Dicke \wh s}

There exist JBD quantum wormhole solutions even in the absence of
matter, in contrast with the situation in \gr.  Since the inclusion
of matter may conceal the actual effect of the JBD scalar, we shall
study the wormhole solutions of a Jordan--Brans--Dicke theory in
vacuum. The matter fields at low energy will feel the existence of
JBD wormholes due to the non trivial coupling of the JBD field to
their masses, absent in general relativity.

Due to the difficulties attending the full field theoretic
expressions that we have to cope with, we shall consider a \mss\
model \cite{HAL} in which the metric and the fields are severely
restricted so that they depend only on a few functions of time
$\tau$. This reduction of the number of \dofs\ will be carried out by
perturbatively expanding the gravitational and the JBD field
variables in harmonics on the three--sphere \cite{HH85}.
This corresponds to an approximately homogeneous
and isotropic \st\ and thus can be described by a
Freedman--Robertson--Walker (FRW) metric plus some perturbations that
represent the inhomogeneities and anisotropies. We shall consider
that all but the first two modes  are in their ground state so that
we can write the \tm, the lapse  and shift functions and the JBD
field as
\be\ba{rl}
\bar \qik(\tau, x^i)\ =&{\displaystyle \frac{2\G}{3\pi}e^{2\al(\tau)}\lp 1+
\sqrt{\frac{2}{3}} a(\tau)Q^{(2)}(x^i)\rp\Om_{ik}(x^i),}\vs\nl
\bar N(\tau, x^i)\ =&{\displaystyle \sqrt\frac{2\G}{3\pi}N_0(\tau)\lp
1+\frac{1}{\sqrt 6}g(\tau)Q^{(2)}(x^i)\rp,}\vs\nl
\bar N_i(\tau, x^i)\ =&{\displaystyle \frac{2\G}{3\pi}e^{\al(\tau)}
\lp\frac{1}{\sqrt 6} k(\tau)P_i^{(2)}(x^i) \rp,}\vs\nl \phi(\tau,
x^i)\ =&{\displaystyle \frac{3\pi}{2\G}
\lp\frac{1}{\pi}\vphi(\tau)+\sqrt{2}f(\tau)Q^{(2)}(x^i)\rp,  }
\ea\eel{QNN}
where $\Om_{ik}$ and $x^i$ are the metric and the coordinates on the
unit three--sphere, $e^\al$ is the usual scale factor in a FRW \st,
$N_0$ and $\vphi$ are the homogeneous modes of the lapse function and
the JBD  field respectively, and $a,\ g,\ k$ and $f$ are the
coefficients of the expansion. $Q^{(2)}$ are the standard scalar
harmonics on the three--sphere and $P^{(2)}_i=\bar{\nabla}_i Q^{(2)}$
are the transverse vector harmonics that correspond
to the first inhomogeneous mode $n=2$. In the expansion of the \tm\
$\bar \qik$, all the other terms, which correspond to the transverse
traceless vector and tensor harmonics,
have not been considered since they will not appear in the action
\cite{HH85}.

The Euclidean action, up to second order in the perturbations, will
then be the sum of the zeroth order action  corresponding to the
homogeneous mode (\ie\ to a FRW\linebreak \st\ plus the homogeneous JBD field),
and the action  associated to the first inhomogeneous mode,
quadratic in the perturbations \cite{HH85}. One can then obtain the
classical Hamiltonian of the theory
\be
\bar H=N_0\lp H_{\mid 0}+H^{(2)}_{\mid 2}+gH^{(2)}_{\mid 1}\rp+kH^{(2)}_{\_1},
\eel{NKG}
where the subindex gives the order in perturbation theory.
Since $N_0,\ k$ and $g$ are not dynamical, they can be
freely varied to yield, by replacing the momenta by
the derivatives with respect to the canonical variables,
the quantum Hamiltonian constraints
\be
H^{(2)}_{\_1}\Psi=0,
\hs\hs
H_{\mid 1}^{(2)}\Psi=0,
\hs\hs
\lp H_{\mid 0}+H_{\mid 2}^{(2)}\rp\Psi=0.
\eel{QHC}
The explicit expressions of these operators can be found in ref.
\cite{HH85}. The first one is the three--\diff\ constraint, which
ensures that the \wf\ is invariant under spatial changes of
coordinates while the last two correspond to the \wdwe. Since the
\diff\ constraint is linear in the gra\-vi\-ta\-tional momentum
$p_a$, we can substitute for $p_a$ and then solve the resultant \wdwe
s in the gauge $a=0$,
\be
\partial_{\vphi}\lp\partial_f-3f\partial_\al\rp\Psi=0,
\eel{wdwe1}
\be
\lp\partial_\al^2-(1-9f^2)\partial_{\vphi}^2-
\partial_f^2-(1-3f^2)e^{4\al}\rp\Psi=0.
\eel{wdwe2}
One can use the solution to these equations as an initial value for
the \diff\ constraint equation and, thus, find the \wf\ for any value
of $a$ \cite{HH85}. Although these equations are formally equivalent
to those of a minimally coupled massless sca\-lar field in
\gr, $\vphi$ is actually the homogeneous scalar component of the
gravitational field, rather than a matter field.

For the \wf\ $\Psi$ to describe a \wh, the JBD field must be
homogeneous at infinity, with an asymptotically zero energy--momentum
tensor \cite{G91}. This means that the perturbations must vanish in
the asymptotic region. Then equations (\ref{wdwe1}, \ref{wdwe2}) reduce to
those of the unperturbed model,
\be
\lp\partial_\al^2-\partial_{\vphi}^2-e^{4\al}\rp\Psi=0,
\eel{wdwu}
whose \wh\ solutions are well known \cite{HP,G91}. In fact one can
construct a complete set of wave functions that generates the whole Hilbert
space of quantum \wh s \cite{G92}.  Far away from the asymptotically
euclidean region, the contribution of the perturbations become
important and, therefore,  the \wh\ \wf\ will have an explicit
non--trivial dependence on them that cannot be gauged away, see
however \cite{LYO}. We have not found explicit analytic expressions
for these  solutions. However, a complete set of solutions must
exist, due to the existence of such a set in the unperturbed model.

Particles may go down the \wh\ from one asymptotically Euclidean
region to another. This can be interpreted from one of these regions
as an effective interaction which modifies the low energy coupling
constants. More precisely, the Green's function between these two
regions can be factorized by inserting a complete set of \wh\
solutions of the \wdwe\ \cite{H88}.  The two--point \wh\ vertex can
then be written as the matrix element between the ordinary flat
space vacuum $\ket{0}$ and the wormhole state $\Psi_{\rm wh}$,
\be
\bra 0\phi(x_1)\phi(x_2)\ket{\Psi_{\rm wh}}=
\int\!\! \cd\bar \qik\p \cd\phi\p\Psi_{\rm wh}(\bar \qik\p,\phi\p)
\int\!\!\cd\bar \gab\cd\phi\ \phi(x_1)\phi(x_2) e^{-\bar I},
\eel{WHV}
where the latter \pin\ is over all \ase\ \fm s $\bar \gab$  and all JBD
fields $\phi$ that have vanishing energy--momentum tensor at
infinity, which match the values $\bar \qik\p$ and $\phi\p$ on a given
\ts. This \pin\ can be evaluated in the saddle point approximation.
{}From the classical equations of the homogeneous and inhomogeneous
parts of the JBD field
\be
\dot{\vphi}+2\vphi\dot \al=0,\hs\hs\dot{f}+3f\dot \al=0,
\eel{CEQ}
respectively, it can be seen that the stationary JBD
configuration falls off like \cite{DOW}
\be
\phi_{\rm hom}(x)\sim\frac{1}{|x-x_0|^2},
\hs\hs
\phi_{\rm inh}(x)\sim\frac{Q^{(2)}(x^i)}{|x-x_0|^3},
\eel{HOM}
for large distance $|x-x_0|$ from the center $x_0$  of the \wh.
Therefore, the two--point \wh\ vertex associated with the
inhomogeneous mode can be found to be
\be
\bra 0\phi(x_1)\phi(x_2)\ket{\Psi_{\rm wh}}=
\mbox{factor}\times\int\!\!d^4x_0\frac{Q^{(2)}\lp (x_1-x_2)^i\rp}
{|x_1-x_0|^3|x_2-x_0|^3},
\eel{FAC}
where an integration over all possible orientations of the \wh\ has
been performed and the prefactor depends on the \wh\ quantum state.
This is precisely the kind of vertex that one would obtain from an
interaction in flat \st\ of the form\linebreak $\displaystyle \int
d^4x_0\lp\partial\phi(x_0)\rp^2$.  Furthermore, in the presence of
low energy matter, wormholes produce an effective interaction which
modifies the masses of elementary particles with an effective
coupling to the scalar field \footnote{In an analogous way, \wh s
will induce an effective potential for the scalar field. However,
observational bounds put very strong constraints on its mass.
{}From now on, we will ignore this
contribution.}. Both effects can be reinterpreted, through a
conformal redefinition of the metric, as giving an effective $\om$
parameter.

\nsection{The effective Brans--Dicke parameter}

As explained in the introduction, string theory predicts an effective
JBD theory of gravity with a particular value, $\om=-1$, of the BD
parameter. One should verify phenomenologically this prediction
\cite{GB92}. Soon after the Planck era, a phase transition due to a
symmetry breaking could be responsible for a short period of extended
inflation \cite{EI}.  In a homogeneous and isotropic \st\, $ds^2 =
-dt^2 + e^{2\al(t)}d\Om^2_3$, the cosmological solutions to the
Lorentzian Jordan--Brans--Dicke equations of motion (which can be
obtained from \form{SPF}), follow the power law evolution \cite{MJ}
\begin{equation}
\label{SOL}
e^{\al(t)} \sim t^{\omega+\frac{1}{2}}, \hspace{1cm} \Phi(t) \sim t^2 .
\end{equation}
A value of $\omega\gg 1$ is necessary for solving the horizon and
flatness problems \cite{EI} \footnote{However, in order not to
disturb the observed isotropy of the cosmic background radiation, we
must also require $\omega\lsim 25$ \cite{EI}.}. String theory
prediction, $\omega=-1$, would give a contraction of the universe,
see eq.~\form{SOL}, instead of an expansion. Furthermore, this value
of $\om$ is also in conflict with primordial nucleosynthesis and
post--Newtonian bounds, $\omega>500$ \cite{PNS,SCT}. We therefore
seem to be very far from the string fundamental description of
gravity.

Nevertheless, one expects that non--perturbative \qg\ wormhole
effects modify the effective value of the Brans--Dicke parameter, as
explained in section 4. We will now follow Coleman's arguments for
the vanishing of the cosmological constant \cite{COL,KSB,H90} in the
context of a JBD theory of gravity. We will obtain that the most
probable value of the effective BD parameter is $\om=\infty$, that
is, we recover general relativity as the low energy effective theory
of gravity.

We cannot observe wormholes, since they do not carry any gauge
charge, energy or momentum \cite{H88}. They are topological quantum
fluctuations of Euclidean spacetime which connect separate
universes. The effect of wormholes can be seen as an insertion in the
field theory path integral of an apparently non--local vertex
operator which can then be rewritten as an effective local
interaction that modifies the low energy coupling constants
\cite{KSB}. These will depend on wormhole configurations labelled by
arbitrary $\al$ parameters. The Euclidean path integral for the whole
universe gives a probability distribution for these $\al$ parameters
and, as a consequence, for the coupling constants of nature
\be
{\cal Z} = \int d\al\ e^{-\frac{1}{2}\al^2} {\cal Z}(\al),
\eel{ZET}
where the weight $\exp(-\al^2/2)$ is determined by the
Hartle--Hawking boundary conditions for the universe \cite{COL},
which were discussed in section 3. If
this probability distribution is strongly peaked at particular values
of these coupling constants, they are a prediction of the theory in
the sense that these will be the value we will observe, with very
high probability.  Coleman assumes that in the path integral one must
sum over all closed manifolds subjected to Hartle--Hawking boundary
conditions. He further assumes that wormholes do not interfere
(dilute wormhole approximation). The partition function ${\cal
Z}(\al)$ is given by the Euclidean \pin\ over all compact manifolds,
containing large smooth compact regions of \st\ disconnected from
each other, except for the existence of \wh s, which can be written
as the exponential of the \pin\ over compact connected manifolds,
\be
{\cal Z}(\al) = \sum_{\rm CM} \int \cd g\cd\Phi\ e^{-I(\al)}
= \exp\sum_{\rm CCM} \int \cd g\cd\Phi\ e^{-I(\al)}.
\eel{CCM}
He then assumes that the most important contribution to the path
integral comes from manifolds with spherical topology. Under those
assumptions one can then evaluate the path integral in the saddle
point approximation.

We will evaluate this path integral in the JBD theory, see eq.
\form{EBD}. As discussed in section 3, Hartle--Hawking boundary
conditions must be imposed in the Jordan frame. We saw that the
Euclidean saddle point corresponds to a compact \st\ only if both the
cosmological constant and the BD parameter are positive.  The
Euclidean action at the saddle point can be computed numerically,
from the eqs. \form{SPF},
with the use of {\em Mathematica} \cite{MAT}. The
resulting effective saddle point action can be written as
\be
I(\al) = - \frac{3\pi}{\G_\al \Lam_\al} f(\om_\al),
\eel{IOM}
where $f(\om_\al)$ is the function shown in fig. 1, whose maximum
value is obtained for $\om_\al = \infty$, where $f(\infty)=1$, for
which we recover the \gr\ result.

The probability distribution for the $\al$ parameters and therefore
for the coupling cons\-tants
\be
{\cal Z}(\al) = \exp\lp\exp \frac{3\pi}{\G_\al\Lam_\al}f(\om_\al)\rp.
\eel{EXP}
is strongly peaked at $\G_\al\Lam_\al = 0$. Since this probability
distribution is not normalizable, a suitable choice of the cutoff is
needed in order to determine its maximum. In the context of \gr,
different cutoffs have been considered \cite{COL,H90,PGW} that give
different results. However, $\G\Lam$ seems to be the most natural
choice for the cutoff since it is the adimensional vacuum energy,
\ie\ the cosmological constant in Planck units, which is observed to
be less than $10^{-120}$. With this choice for the cutoff, the
probability distribution \form{EXP} acquires its maximum at $\om_\al
= \infty$.  Therefore, general relativity is a prediction of this
scenario based on non--perturbative \qg\ effects. This means that
there can be no effective dynamical scalar field coupled
to energy or matter. In particular, even in the case that different
kinds of matter had different couplings to the scalar field
\cite{DGG,CQG,GB92}, the weak equivalence principle will be effectively
recovered due to wormholes.  Note that this result can be generalized
to any scalar--tensor theory with an arbitrary coupling $\om(\Phi)$.

It is important to understand the physical meaning of this
prediction. The usual\linebreak interpretation \cite{COL} is that Coleman's
mechanism only ensures that the bottom line cosmological constant is
zero. This does not exclude an inflationary universe with a non--zero
false vacuum energy due a certain phase transition. It just says
that, whatever the effective potential for inflation is, the true
minimum is at zero cosmological constant. However, it does preclude
extended inflation since wormholes drive the BD parameter to infinity
at all times and thus ``freeze out" the evolution of the scalar field.

\nnsection{Acknowledgments}

We would like to thank Pedro Gonz\'alez--D\'{\i}az, Guillermo Mena
Marug\'an and Mariano Quir\'os for a careful reading of the
manuscript and valuable discussions. We also thank Andrei Linde for
clarifying our conclusions on extended inflation.

\newpage
%\section*{Figure Captions}
.\vspace{16cm}
\begin{description}

\item[Fig.1]
Plot of the function $f(\om)$ which determines the saddle point
action \form{IOM}. The $\om<0$ region has been excluded since it
corresponds to non--compact saddle points. This function acquires its
maximum at $\om=\infty$, where $f(\infty)=1$, which corresponds to
the \gr\ result.

\end{description}

\end{document}